\documentclass[a4paper,11pt]{article}
\usepackage{pos}
\usepackage{enumitem}
\usepackage{tikz}
\usepackage{tikz-feynman}

\title{Quantum properties of $U(1)$-like gauge theory on $\kappa$-Minkowski}

\author*[a]{Kilian Hersent}

\affiliation[a]{IJCLab, Universit\'{e} Paris-Saclay, CNRS/IN2P3,\\%
  91405 Orsay, France}

\emailAdd{kilian.hersent@universite-paris-saclay.fr}

\abstract{
In the $5$-dimensional twisted $U(1)$-like gauge theory on $\kappa$-Minkowski, the one-loop one-point (tadpole) function was computed in \cite{Hersent_2022b}. This article summarizes the construction of such a gauge theory and discusses the non-vanishing of the tadpole. 
}

\FullConference{%
  Corfu Summer Institute 2022 "School and Workshops on Elementary Particle Physics and Gravity",\\
  28 August - 1 October, 2022\\
  Corfu, Greece}

\DeclareMathOperator{\actr}{\triangleleft}
\DeclareMathOperator{\actl}{\triangleright}
\DeclareMathOperator{\bicros}{\actl\!\!\!\blacktriangleleft}
\DeclareMathOperator{\convp}{\hat{\circ}}

\newcommand{\td}{\mathrm{d}}

\begin{document}
\maketitle

\paragraph{}
The $\kappa$-Minkowski space correspond to a (quantum) deformation of the Minkowski space and is thought to light some quantum gravity effects up, at least in some regime. This thought comes from several arguments. 

First, $\kappa$-Minkowski is build as the space having the symmetries of $\kappa$-Poincar\'{e}, a deformation of the Poincar\'{e} algebra. Therefore, they both intrinsically encode an energy scale $\kappa$, thought to be of the order of the Planck mass, at which they are relevant. Their low energy limit, that is when $\kappa \to + \infty$, corresponds to the usual Minkowski space and its usual Poincar\'{e} symmetries. A field theory on $\kappa$-Minkowski could then by construction contain both the low-energy physics and corrections due to $\kappa$ deformation.

Second, the $\kappa$-Poincar\'{e} algebra realises a Doubly Special Relativity. Indeed, in such a space, the composition of momenta is also deformed. Therefore, changing from one frame to another by such composition gives a conservation of the speed of light but also of the energy scale $\kappa$. The Doubly Special Relativity framework thus gives two upper bounds on physical speed and energy. This framework has been shown to trigger some phenomenological effects such as time-delay observation from highly energetic astrophysical sources \cite{Amelino_Camelia_1998}. For a complete review on phenomenology of quantum gravity see \cite{Addazi_2022}.

\paragraph{}
For a physical insight of field theories on $\kappa$-Minkowski, one could consider a gauge theory. For a complete review about gauge theory on quantum spaces, see \cite{Hersent_2022c}. In this paper, the simplest gauge theory is considered and turns out to be a deformed $U(1)$ gauge theory. The one-loop one-point function, also called the tadpole, was computed and turns out to be non-zero \cite{Hersent_2022b} and even gauge dependant. This may either point out some instabilities in this theory, or some deeper results and prospects concerning noncommutative gauge theories.

The section \ref{sec:kM} introduces briefly the $\kappa$-Minkowski space. The gauge theory on $\kappa$-Minkowski is constructed in section \ref{sec:gt} and the computation of the tadpole is discussed in section \ref{sec:opf}.

\section{Introducing \texorpdfstring{$\kappa$}{kappa}-Minkowski}
\label{sec:kM}
\paragraph{}
We present here a brief construction of the $\kappa$-Minkowski space $\mathcal{M}_\kappa$. For a more complete review on $\kappa$-Minkowski history and phenomenology, see \cite{Lukierski_2017}. In order to build the $\kappa$-Minkowski space, one must first study the $\kappa$-Poincar\'{e} algebra.

\subsection{The symmetries of \texorpdfstring{$\kappa$}{kappa}-Minkowski}
\label{subsec:kM_kP}
\paragraph{}
The derivation of the $\kappa$-Poincar\'{e} algebra $\mathcal{P}_\kappa$ was first done in \cite{Lukierski_1991}. It is defined as Hopf algebra \cite{Klimyk_1997} having as generators $(P_\mu)_{\mu = 0, \dots, d}$, $(M_j)_{j = 1, \dots, d}$ and $(N_j)_{j = 1, \dots, d}$, where $d$ is the space dimension. They are respectively called the deformed translations, the deformed rotations and the deformed boosts, as they reduces to the usual translations, rotations and boosts of the Poincar\'{e} algebra in the limit $\kappa \to +\infty$.

In \cite{Majid_1994}, the $P_0$ generator is changed for the generator $\mathcal{E} = e^{- P_0/\kappa}$. We will follow this choice throughout this paper. The latter authors showed the $\kappa$-Poincar\'{e} can be defined through the bicrossproduct $\mathcal{P}_\kappa = \mathcal{T}_\kappa \bicros U\mathfrak{so}(1,d)$, where $\mathcal{T}_\kappa$ is often called the deformed translation algebra and is generated by the $P_\mu$'s and $U\mathfrak{so}(1,d)$ corresponds to the universal enveloping algebra of the Lie algebra of rotation and boosts $\mathfrak{so}(1,d)$.

\paragraph{}
$\kappa$-Minkowski is then defined as the Hopf dual of $\mathcal{T}_\kappa$ \cite{Majid_1994}, which implies that $\mathcal{M}_\kappa$ is generated by $(x^\mu)_{\mu=0, \dots, d}$ and satisfies
\begin{align}
	[x^0, x^j] = \frac{i}{\kappa} x^j, &&
	[x^j, x^k] = 0.
	\label{eq:kM_kP_kM_alg}
\end{align}
Then, the dual and the bicrossproduct structures allow to define the action of $\mathcal{P}_\kappa$ on $\mathcal{M}_\kappa$. For $\mathcal{T}_\kappa$, it writes for any $f \in \mathcal{M}_\kappa$,
\begin{align}
	(P_\mu \actl f)(x) = - i \partial_\mu f(x), &&
	(\mathcal{E} \actl f)(x) = f( x^0 + \frac{i}{\kappa}, x^j ).
	\label{eq:kM_kP_act}
\end{align}

For simplicity, we avoid the full Hopf algebra structure of $\mathcal{P}_\kappa$, that can be found in \cite{Hersent_2022c}. Still, the coalgebra sector of $\mathcal{T}_\kappa$ is of interest. Indeed, it writes
\begin{align}
	\Delta P_0 = P_0 \otimes 1 + 1 \otimes P_0, &&
	\Delta P_j = P_j \otimes 1 + \mathcal{E} \otimes P_j, &&
	\Delta \mathcal{E} = \mathcal{E \otimes E}.
	\label{eq:kM_kP_kP_coalg}
\end{align}
First, this relation shows that $\mathcal{T}_\kappa$ is not cocommutative which, by duality, imposes that $\mathcal{M}_\kappa$ is noncommutative. Second, by the action \eqref{eq:kM_kP_act} and the comodule algebra structure, one obtains that some generators must follow a deformed Leibniz rule. Explicitly, if we consider a star-product $\star$ on $\mathcal{M}_\kappa$, writing $X_j = P_j$ and $X_0 = \kappa(1-\mathcal{E})$, one has
\begin{align}
	X_\mu \actl (f \star g)
	= (X_\mu \actl f) \star g + (\mathcal{E} \actl f) \star (X_\mu \actl g),
	\label{eq:kM_kP_tw_Lr}
\end{align}
for any $f,g \in \mathcal{M}_\kappa$. The equation \eqref{eq:kM_kP_tw_Lr} correspond to a twisted Leibniz rule and so defines twisted derivations $X_\mu$. For simplicity, we drop the action symbol $\actl$ in the following.

\subsection{The star-product structure}
\label{subsec:kM_sp}
\paragraph{}
We follow the procedure of \cite{Hersent_2022c} to build the star product leading to the $\kappa$-Minkowski star-product of \cite{Durhuus_2013, Poulain_2018}, where more mathematical insight can be found. 

Considering the Lie algebra \eqref{eq:kM_kP_kM_alg}, the corresponding Lie group is the affine group $\mathcal{G} = \mathbb{R} \ltimes \mathbb{R}^d$. Its right invariant Haar measure can be shown to be the usual Lebesgue measure and will therefore be the one considered. From that, one defines the convolution product $\convp$ and the involution ${}^*$ of $L^1(\mathcal{G})$. Then, one defines the Weyl quantization map $Q = \pi \circ \mathcal{F}$, where $\pi$ is a bounded non-degenerate ${}^*$-representation of $L^1(\mathcal{G})$ and $\mathcal{F}$ is the Fourier transform. Finally, the star-product $\star$ and involution ${}^\dagger$ of $\kappa$-Minkowski are defined as $f \star g = Q^{-1}(Q(f) \convp Q(g))$ and $f^\dagger = Q^{-1}( Q(f)^*)$, which gives
\begin{subequations}
\begin{align}
	(f \star g)(x^0, x^j)
	&= \int \frac{\td p_0}{2\pi} \td y^0\ e^{-i p_0 y^0} f(x^0 + y^0, x^j) g(x^0, e^{-p_0/\kappa} x^j),
	\label{eq:kM_sp_sp} \\
	f^\dagger(x^0, x^j)
	&= \int \frac{\td p_0}{2\pi} \td y^0\ e^{-i p_0 y^0} \overline{f}(x^0 + y^0, e^{-p_0/\kappa} x^j),
	\label{eq:kM_sp_inv}
\end{align}
	\label{eq:kM_sp}
\end{subequations}
where $\overline{f}$ is the complex conjugate of $f$.

\paragraph{}
Given the star-product \eqref{eq:kM_sp_sp}, the Lebesgue integral defines a twisted trace, that is for any $f,g \in \mathcal{M}_\kappa$,
\begin{align}
	\int \td^{d+1}x\ f \star g
	= \int \td^{d+1}x\ \mathcal{E}^d(g) \star f.
	\label{eq:kM_sp_ttr}
\end{align}

\section{Gauge theory on \texorpdfstring{$\kappa$}{kappa}-Minkowski: a twisted approach}
\label{sec:gt}
\paragraph{}
We consider here derivation based gauge theory \cite{Dubois-Violette_2001}. This first requires to take a $\mathcal{M}_\kappa$-module $\mathbb{E}$, that is a linear space with an action $\actr : \mathbb{E} \otimes \mathcal{M}_\kappa \to \mathbb{E}$ satisfying $(m \actr f) \actr g = m \actr (f \star g)$ for any $m \in \mathbb{E}$. The simplest relevant choice for this module is to take it as a copy of our algebra. Explicitly, $\mathbb{E} = \mathcal{M}_\kappa$ and $\actr = \star$. This choice will be kept throughout the paper.

\subsection{Untwisted gauge theory}
\label{subsec:gt_not}
\paragraph{}
Building a gauge theory on $\kappa$-Minkowski has long been considered a complicated task due to the non-cyclic property of the trace \eqref{eq:kM_sp_ttr}. Indeed, if we consider a usual gauge theory on a quantum space, the gauge group is the set of elements satisfying $g \star g^\dagger = 1$ and the deformed field strength $F$ transforms as $F^g = g^\dagger \star F \star g$. If we further choose the action to be the straightforward generalisation of the usual $U(1)$ action $S = \int \td^{d+1}x\ F^\dagger \star F$, then the gauged transformed action writes
\begin{align}
	S^g
	= \int \td^{d+1}x\ (F^g)^\dagger \star F^g
	= \int \td^{d+1}x\ \mathcal{E}^d(g) \star g^\dagger \star F^\dagger \star F
	\neq S,
	\label{eq:gt_not_act}
\end{align}
where we used \eqref{eq:kM_sp_ttr}. Therefore, $S$ is not gauge invariant because of the twisted property of the trace. In section \ref{subsec:gt_der}, a twisted derivation based gauge theory was used to bypass this issue.

\paragraph{}
Apart from the twisted framework developed in this paper, there has been two previous attempts in building a gauge theory: \cite{Dimitrijevic_2004a, Dimitrijevic_2004b, Dimitrijevic_2004c, Dimitrijevic_2004d, Dimitrijevic_2005} and \cite{Dimitrijevic_2011, Dimitrijevic_2014}. However, these method relies on intractable star-products so that they can only produce first order corrections. One can also mention another recent attempt based on Poisson gauge theory \cite{Kupriyanov_2020, Kupriyanov_2021}. For a more complete discussion on these gauge theory see \cite{Hersent_2022c}.

\subsection{Twisting the derivations}
\label{subsec:gt_der}
\paragraph{}
The non-cyclicity of the trace \eqref{eq:kM_sp_ttr}, hampering from having a straightforward gauge invariant action, is a first hint that our theory must be twisted somehow. Furthermore, the twists considered will most probably be powers of $\mathcal{E}$, since the twist of \eqref{eq:kM_sp_ttr} is $\mathcal{E}^d$. This choice of twist is also convenient since, at the low energy limit $\kappa \to +\infty$, it vanishes, \textit{i.e.}\ $\mathcal{E} = 1$. The low energy limit will then be by essence untwisted. Finally, we have natural candidates for twisted derivations, which are the $X_\mu$'s from \eqref{eq:kM_kP_tw_Lr}.

To build our gauge theory, we will then start from twisted derivations. This gauge theory was first considered in \cite{Mathieu_2020a, Mathieu_2020b}. One should note that, the twist $\mathcal{E}^d$ of \eqref{eq:kM_sp_ttr} contains the space dimension $d$ so that a gauge invariant action may require $d$ to take some specific value.

\paragraph{}
When considering twisted derivations, one \textit{has to} consider twisted connection, curvature and gauge transformations to ensure their algebraic compatibility with the twisted derivations. Then, for our module $\mathbb{E} = \mathcal{M}_\kappa$, the twisted connection $\nabla_\mu : \mathcal{M}_\kappa \to \mathcal{M}_\kappa$ must satisfy, for any $f, g \in \mathcal{M}_\kappa$,
\begin{align}
	\nabla_\mu (f \star g)
	= \nabla_\mu(f) \star g + \mathcal{E}(f) \star X_\mu(g),
	\label{eq:gt_der_conn}
\end{align}
where $\nabla_\mu = \nabla_{X_\mu}$. One then defines the deformed gauge potential as $A_\mu = \nabla_\mu(1)$.

The deformed field strength then expresses as
\begin{align}
	F_{\mu\nu}
	= X_\mu(A_\nu) - X_\nu(A_\mu) + \mathcal{E}(A_\mu) \star A_\nu - \mathcal{E}(A_\nu) \star A_\mu,
	\label{eq:gt_der_curv}
\end{align}
where the twist $\mathcal{E}$ appears again.

\paragraph{}
The gauge group derivation appears to be untwisted and correspond to a deformed $U(1)$ group
\begin{align}
	\mathcal{U}(1)
	= \{g \in \mathcal{M}_\kappa,\
	g^\dagger \star g = g \star g^\dagger = 1 \}.
	\label{eq:gt_der_gg}
\end{align}
However, the gauge transformation of the potential and the field strength are twisted and writes
\begin{align}
	A_\mu^g
	&= \mathcal{E}(g^\dagger) \star A_\mu \star g + \mathcal{E}(g^\dagger) \star X_\mu(g), &
	F_{\mu\nu}^g
	&= \mathcal{E}^2(g^\dagger) \star F_{\mu\nu} \star g.
	\label{eq:gt_der_gt}
\end{align}

\subsection{The gauge invariant action}
\label{subsec:gt_act}
\paragraph{}
Consider now the previous action $S = \int \td^{d+1}x\ F^\dagger \star F$. Doing again the computation \eqref{eq:gt_not_act} with the transformation \eqref{eq:gt_der_gt}, one obtains that $S$ is gauge invariant if $d+1=5$. Therefore, the gauge invariant action writes
\begin{align}
	S = \int \td^5x\ F^\dagger \star F.
	\label{eq:gt_act_gia}
\end{align}
This action is then invariant under $\mathcal{U}(1)$ \eqref{eq:gt_der_gg} transformations. Moreover, one can show that it is also $\kappa$-Poincar\'{e} invariant. Finally, its low energy limit $\kappa \to +\infty$ coincide with the standard Abelian Yang-Mills action in $5$ dimensions.

First, note that this dimension constraint is rather robust since a twisted action $\actr$ was considered, but leaves the constraint untouched \cite{Hersent_2022a}. Explicitly, by setting $m \actr f = m \star \sigma(f)$ with $\sigma \in \mathrm{Aut}(\mathcal{M}_\kappa)$ an automorphism of $\kappa$-Minkowski, $d+1=5$ is still necessary for $S$ to be gauge invariant. Second, the phenomenological consequences of the extra-dimension has been discussed in \cite{Mathieu_2020b}, where a compactification scheme is considered.

\section{The one-loop one-point function}
\label{sec:opf}
\paragraph{}
From \eqref{eq:gt_der_curv}, one can put the action \eqref{eq:gt_act_gia} under the form
\begin{align}
	S = \int \td^5x\
	K^{\mu\nu} A_\mu A_\nu
	+ V_{(3)}^{\mu\nu\rho} A_\mu A_\nu A_\rho
	+ V_{(4)}^{\mu\nu\rho\sigma} A_\mu A_\nu A_\rho A_\sigma
	\label{eq:gt_opf_action}
\end{align}
where $A$ was assumed to be real valued, \textit{i.e.}\ $\overline{A} = A$. One can compute the one-loop one-point function via the Fadeev-Popov procedure \cite{Hersent_2022b} with a BRST gauge fixing \cite{Mathieu_2021}. Two different gauge fixing were considered that writes $X^\mu A_\mu = 0$ and $A_0 = \lambda$. The first one correspond to a deformation of the Lorenz gauge fixing and the second one is a parametrized temporal gauge fixing, the temporal gauge being recovered when $\lambda \to 0$. One obtains that
\begin{align}
	\feynmandiagram [medium, baseline=(a.base), horizontal=a to b] {
  		a -- [photon] b [blob],
	}; \
	&=
	\left\{ \begin{matrix}
		\int \td^5x\; \mathcal{I}(\kappa) A_0(x)
		& (X^\mu A_\mu = 0) \\
		\Gamma^{\mathrm{gh}}_1(A_0) + \lambda \int \td^5x\; \mathcal{J}(\kappa) A_0(x)
		& (A_0 = \lambda)
	\end{matrix} \right. ,
	\label{eq:gt_opf_tadpole}
\end{align}
where $\mathcal{I}$ and $\mathcal{J}$ are gauge dependant divergent integral (which must be regularized), that vanish when $\kappa \to \infty$ and $\Gamma_1^{\mathrm{gh}}$ is the ghost contribution to the tadpole.

\paragraph{}
The fact that \eqref{eq:gt_opf_tadpole} is non zero implies that the classical vacuum is not stable against quantum fluctuation. In other words, $A_\mu$ has a non-zero vacuum expectation value. Several comments are in order.

First, the condition $\overline{A} = A$ can be relaxed and has been shown to be non-vanishing also. Adding matter in the action \eqref{eq:gt_opf_action}, as detailed in \cite{Mathieu_2020b}, gives a zero contribution to the tadpole.

Second, in the low energy limit $(\kappa \to + \infty)$ both expressions of \eqref{eq:gt_opf_tadpole} vanishes. Therefore, the good commutative limit is observed. Recall that such a commutative limit correspond to $5$-dimensional $U(1)$ gauge theory.

Finally, the result \eqref{eq:gt_opf_tadpole} is explicitly gauge dependant. As we started from a gauge invariant action \eqref{eq:gt_act_gia}, the gauge symmetry has been broken in the procedure. Note that in the temporal gauge $(\lambda \to 0)$ one has a zero tadpole \eqref{eq:gt_opf_tadpole}, since the ghost decouples. However, this is purely a gauge artefact.

\paragraph{}
Still, several arguments put forward that our theory is not to be thrown away so quickly. 

Indeed, a non-zero tadpole has been experienced in other quantum spaces like $2$-dimensional Moyal \cite{Martinetti_2013} or $\mathbb{R}^3_\lambda$ \cite{Gere_2014}, with the same computation procedure. This could imply that one cannot apply the usual Feynman path integral method \cite{Camacho_2005} and/or BRST liturgie in a noncommutative context.

Moreover, the notion of vacuum of $\kappa$-Minkowski has not reach consensus. First, considering $\kappa$-Poincar\'e as the space of symmetry, one get rid of Poincar\'{e} symmetry and so looses the notion of particle as irreducible representation of the little group. This issue is quite known as it appears in quantum field theory in curved spacetime, see \cite{Ford_2021} for a review. Some studies shows that the representation theory of $\kappa$-Poincar\'{e} and its little group are quite close to the undeformed one \cite{Ruegg_1994, Rouhani_1994, Kosinski_1995, Kosinski_2001}. Still, the definition of $\kappa$-Minkowski vacuum and its energy, as studied in \cite{Cougo_Pinto_2004, Kim_2008, Harikumar_2013, Juric_2019}, has several definitions and has even shown some pathological behaviour. This issue can be traced back to the fact that the various definitions of the Casimir operator are coordinate dependant. Therefore, the vacuum expectation value of $A$, as computed in \eqref{eq:gt_opf_tadpole}, is non-zero as it possibly may not be expressed in the physical vacuum.

Finally, some studies on the Moyal space are considering an alternative field as $A$ to encode the photon \cite{Aoki_2000, Madore_2000, Bichl_2002, Wulkenhaar_2002, Steinacker_2007, De_Goursac_2007, Grosse_2008}. The latter is called the ``covariant coordinate'' and is expressed as $\mathcal{A}_\mu = A_\mu - i \Theta^{-1}_{\mu\nu}x^\nu$, with $[x^\mu, x^\nu] = \Theta^{\mu\nu}$. From \cite{Madore_2000}, it appears that the covariant coordinate of $\kappa$-Minkowski may not be computable. However, the fact that \eqref{eq:gt_opf_tadpole} is non vanishing may hint the possibility that $A$ is not the physical photon of the theory.

\section*{Acknowledgement}
\paragraph{}
The author thanks S.\ Clery, M.\ Dimitrievi\'{c}-Ciri\'{c}, C.\ P.\ Martin, A.\ Sitarz, P.\ Vitale and J.-C.\ Wallet for enlightening discussions.


\end{document}